\documentclass[preprint,aps,showpacs]{revtex4}
\usepackage[pdftex]{graphicx}
\usepackage{amsmath}
\usepackage{amsxtra}
\usepackage{amstext}
\usepackage{amssymb}
\usepackage{latexsym}
\usepackage{dsfont}
\usepackage{slashed}

\newcommand{\be}{\begin{equation}}
\newcommand{\ee}{\end{equation}}
\newcommand{\bn}{\begin{eqnarray}}
\newcommand{\en}{\end{eqnarray}}

\begin{document}

\title{How many scalar fields there are and how  are they related 
to fermions and weak bosons in the {\em spin-charge-family} theory?}

\author{Norma Susana Manko\v c  
Bor\v stnik}
\affiliation{Department of Physics, FMF, University of Ljubljana,
Jadranska 19, 1000 Ljubljana, Slovenia}


\begin{abstract}
The {\em  spin-charge-family} theory~\cite{norma,pikanorma,NF,gmdn,gn} offers a possible explanation for  
the assumptions of the {\it standard model}, interpreting the {\em standard model} as its low 
energy effective manifestation~\cite{NF}.  The {\it standard model} Higgs 
and Yukawa couplings are explained as an effective replacement for several scalar fields, all of 
bosonic (adjoint) representations with respect to all the charge groups, with the family groups 
included. Assuming 
the Lagrange function for all scalar fields to be of the renormalizable kind, properties of the scalar fields 
on the tree level are discussed. Free scalar fields (mass eigen states) differ from either those, 
which couple to $Z_m$,  or to $W^{\pm}_{m}$ or to each family member of each of the four families, 
which further differ among themselves. Consequently the {\em  spin-charge-family} theory predictions 
differ from those of the {\it standard model}.  
\end{abstract}

\keywords{Unifying theories, Origin of families, Properties of scalar fields, The fourth family, 
Fermion properties, Boson masses, Flavour symmetry}

\pacs{14.60.Pq, 12.15.Ff, 12.60.-i}
\maketitle

\section{Introduction } 
\label{introduction}

The {\it standard model} assumes one scalar field, the Higgs, with the charges in the fundamental 
representation of the charge groups and the Yukawa couplings. The Higgs, forming a kind of the 
"scalar Majorana", and interacting with the weak and hyper gauge bosons, determines masses of 
the weak bosons and, "dressing" right handed fermions with the needed weak and hyper charges, 
determines, together with the Yukawa couplings, also the masses of the so far observed families 
of fermions.

The question is: Where do the Higgs together with the Yukawa couplings  of the {\it standard model} 
originate from?  
Is the Higgs really a scalar field with the fermionic quantum 
numbers in the charge sector, or it is just, so far extremely efficient, effective  representation 
for several scalar fields which manifest as the Higgs and the  Yukawa couplings?

To answer any question about the origin of the Higgs and the Yukawa couplings one first needs the answer 
to the question:
Where do families originate? And correspondingly: How many 
families do we have at all?

Effective interactions can have in physics many times quite unexpected shapes and yet 
can be very useful (as is the case, for example, with by the experiments suggested spin-spin 
interactions in several models describing (anti)feroelectric and (anti)feromagnetic 
materials and also other properties, 
where the interaction of the electromagnetic origin among many electrons and nuclei involved 
can effectively be expressed by a kind of the "spin-spin" interaction).

I am proposing the theory~\cite{norma,pikanorma,NF,gmdn,gn,AN,HN},  the {\it  spin-charge-family} theory, 
which does offer the explanation for the origin of families, of  vector gauge fields and of  
several scalar fields: A simple starting action  
at higher dimensions determines at low energies properties of families of fermions, of the known vector 
gauge fields and of several scalar fields. Vector and scalar fields, which originate in the spin 
connections and vielbeins at higher dimensions, have correspondingly 
the charges in the adjoint representations with respect to all charge groups. 
Families appear in the theory  due to the fact that there are  two 
kinds of gamma matrices (two kinds of the Clifford algebra objects, only two): 
i. The one used by Dirac to describe the spin of fermions (spinors).  ii. The second 
one used in this theory to  explain the origin of  families~\footnote{ 
More about the two kinds of the Clifford algebra objects can be found in the 
refs.~\cite{norma,pikanorma,hn0203,NF}. The appendix~\ref{scalartriplets} 
gives a short overview.}. The theory predicts before the electroweak break four 
families of ($u^i$ and $d^i\,$, $i\in \{1,2,3,4\}$) quarks and 
($e^i$ and $\nu^i$) leptons,  left  handed weak charged and right handed weak chargeless, which when 
coupling to massive scalar fields with non zero vacuum expectation values and to gauge fields, 
become massive.

In this paper the scalar fields  and their  taking care of the masses of quarks and leptons 
and of the weak bosons on the tree level are studied and the predictions discussed. 
Although the {\it  spin-charge-family} theory 
still requires (many) additional studies to be proved -- or disproved -- that it is the right step 
beyond the {\it standard model}, yet the work done so far~\cite{norma,pikanorma,NF,gmdn,gn,HN,NHD,AN}  
gives a hope.   

Keeping the symmetries of  four  
massless families (the spins and charges of quarks and leptons, left and right handed)  
and of the massless gauge fields and the scalar fields as they follow from the {\it spin-charge-family} 
theory before the electroweak break and letting the scalar fields to 
gain nonzero vacuum expectation values, the tree level contributions  strongly relate 
properties of family members  and also of the gauge weak fields. It is a hope, supported by the calculations 
done so far~\cite{AN,gmdn}, that loop corrections (in all orders) lead to the observed 
properties of fermions.  

In this paper I assume (not  derive from the {\it  spin-charge-family} theory)  the Lagrange 
function for the scalar fields, which cause the electroweak break, and their couplings 
 to  gauge bosons~(Eq.(\ref{scalarLagrange0})) 
 so that the  theory is renormalizable. 
 This assumption 
 is made only to manifest that 
 the existence of several scalars might strongly influence the experiments, which search for the Higgs.

After the electroweak break the effective Lagrange density for the four families of fermions  looks in the 
{\it spin-charge-family} theory~\cite{NF} as 
 \begin{eqnarray}
 {\mathcal L}_f &=&  \bar{\psi}\, (\gamma^{m} \, p_{0m} - \,M) \psi\, , \nonumber\\
          p_{0m}&=& p_{m} - \{ g^{Y}\,\cos \theta  \,Q\,A_{m} +  g^{1} \cos \theta \,Q'\, Z^{Q'}_{m} +
          \frac{g^{1}}{\sqrt{2}}\, (\tau^{1+} \,W^{1+}_{m} + \tau^{1-} \,W^{1-}_{m}) + 
          g^{Y'} \cos \theta^{'} \,Y'\, A^{Y'}_{m}
          \,\} \nonumber\\
           \bar{\psi}\,  M \, \psi &=&  \bar{\psi}\, \gamma^{s} \, p_{0s} \psi \,,\quad   
           s\in\{7,8\}\,, \nonumber\\
           %
          p_{0s}&=& p_{s} - \{ \tilde{g}^{\tilde{N}_L} \,\vec{\tilde{N}}_L\, \vec{\tilde{A}}^{\tilde{N}_L}_{s} +
 		  	                \tilde{g}^{\tilde{Q}'} \, \tilde{Q}'\,
 		  	                \tilde{A}^{\tilde{Q}'}_{s}
 		  	       	          + \frac{\tilde{g}^{1}}{\sqrt{2}}\, (\tilde{\tau}^{1+}
 		  	       	          \,\tilde{A}^{1+}_{s}
 	          + \tilde{\tau}^{1-}\,  \tilde{A}^{1-}_{s})\nonumber\\ 	
 	          &+&  g^{Y}_{1}\, \cos \vartheta_1 \, Q\, A^{Q}_{s} +  g^{1}_{1}\, 
 	          \cos \vartheta_1 \,Q'\, Z^{Q'}_{s} +
          \frac{g^{1}}{\sqrt{2}}\, (\tau^{1+} W^{1+}_{s} + \tau^{1-} W^{1-}_{s})\, \nonumber\\
               &+&g^{2} \cos \vartheta'_2\, Y'\, A^{Y'}_{s}\,\}\,,\quad 
 	           Q= \tau^{13} + Y\,,\quad Q'_{1}= (\tau^{13} - \tan^2 \vartheta_1 \,Y)\,,
 	           \quad \frac{g^{Y}_{1}}{g^{1}_{1}}= \tan \vartheta_1 \,.
 \label{factionI}
 \end{eqnarray}
$Q$ and $Q'$  are the {\it standard model} like  charges ($ \vartheta_{1}$ does not need to be $\theta$),
 while $Y'$ is the additional one~\cite{NF},  appearing in the {\it spin-charge-family} theory 
after the break which is followed by the electroweak break. 

Taking into account that  $\stackrel{78}{(\pm )}= \frac{1}{2}(\gamma^{7}\pm \gamma^{8}) $, the mass term 
$\bar{\psi}\,  M \, \psi $ of the Lagrange function ${\mathcal L}_f $ can be 
rewritten as follows
\begin{eqnarray}
\label{pm}
\bar{\psi}\,  M \, \psi &=&  \bar{\psi}\,\stackrel{78}{(-)} \,p_{0\,-} +  
\stackrel{78}{(+)}\,p_{0\,+} \psi \,,\quad p_{0\,\mp} = (p_{0\,7}\pm ip_{0\,8})\,.
\end{eqnarray}
The mass term $\bar{\psi}\,  M \, \psi $  determines the tree level mass matrices of quarks and 
leptons and correspondingly the masses, the Yukawa couplings and the mixing matrices  
for the four families after the loop corrections are taken into account. 
The  scalar fields -- the triplets ($\vec{\tilde{A}}^{\tilde{N}_L}_{\mp}$, 
$\vec{\tilde{A}}^{\tilde{1}}_{\mp}$,
$\vec{A}^{1}_{\mp}$, $\vec{A}^{o}_{\mp}$) and the singlets  
($A^{Y'}_{\mp}$, $A^{Y}_{\mp}$) 
-- are with their vacuum expectation values responsible for the 
appearance of the fermionic masses~\footnote{The two triplets ($\vec{\tilde{A}}^{\tilde{N}_L}_{\mp}$, 
$\vec{\tilde{A}}^{\tilde{1}}_{\mp}$) and the singlets ($A^{Q}_{\mp}$, $A^{Q'}_{\mp}$, $A^{Y'}_{\mp}$) 
appear in the mass  
term (Eq.(\ref{factionI})) contributing to masses of fermions since they are assumed to have nonzero vacuum 
expectation values. Since  $\tau^{1i}$ 
$\psi^{\alpha}_{R}=0$, the 
$\vec{A}^{1}_{\mp}$  and   $\vec{A}^{o}_{\mp}$ triplets
do not contribute to the masses of fermions on the tree level.}. 

The existence of the two scalar 
triplets, $\vec{A}^{1}_{\mp}$ and $\vec{A}^{o}_{\mp}$, with the properties 
\begin{eqnarray}
\label{tripletvacuu}
Y\, \vec{A}^{1}_{\mp}&=& - \vec{A}^{1}_{\mp}\,,\quad T^{13}\, \vec{A}^{1}_{\mp}= (A^{1 1}_{\mp}, 0,-A^{1 2}_{\mp})\,,
\quad <\vec{A}^{1}_{\mp}>= (v_{11\,\mp},0,0)\,,\nonumber\\
Y\, \vec{A}^{o}_{\mp}&=& 0\,,\quad T^{13}\, \vec{A}^{o\,1}_{\mp}= (A^{o\,1}_{\mp}, 0,-A^{o\,2}_{\mp})\,,
\quad <\vec{A}^{o}_{\mp}>= (0,v_{o\,3\,\mp},0)\,,
\end{eqnarray}
where $<\vec{A}^{1}_{\mp}>$ and $<\vec{A}^{o}_{\mp}>$ denote the vacuum expectation values, are assumed. 
Both contribute to the masses of the weak bosons on the tree level. 
Although all the scalar fields have in the {\it spin-charge-family} theory 
charges in the adjoint representation, it can not be expected that a triplet is a good replacement for  
several scalar fields -- vielbeins and spin connections of the two kinds -- involved in the electroweak break.
I also take into account 
only the tree level contributions, although it might turn out that the loop corrections are very important 
and are needed probably in all orders to reproduce the measured properties of fermions and weak bosons.

The tree level mass term $\bar{\psi}\,  M \, \psi $ distinguishes among the members of any of the 
four families -- due to the operators $Q$ and $Q'$  and due to the operators $\stackrel{78}{(\mp )} $, 
which transform the weak and hyper charges of $u^{i}_R$-quarks and $\nu^{i}_R$
-leptons ($\stackrel{78}{(-)}$) into those of the corresponding left handed ones and the weak and hyper charges 
of $d^{i}_R$-quarks and $e^{i}_R$-leptons ($\stackrel{78}{(+)}$) into those of their 
left handed ones~\cite{NF}. The reader can find a short explanation in appendix~\ref{cliffordfamilies}. 
The fields  $\tilde{A}^{\tilde{N}_L i}_{\mp}$ and  $\tilde{A}^{1i}_{\mp}$,  the two kinds of triplets,
couple before the electroweak break  to the  family members due to the operators $\vec{\tilde{\tau}}^{1}$ and 
$\vec{\tilde{N}}_L$ as  
demonstrated in the diagram below
\begin{equation}
\label{diagramNtau}
      \stackrel{\stackrel{\vec{\tilde{N}}_{L}}{\leftrightarrow} }{\begin{pmatrix} III & IV \\
I & II  \end{pmatrix}} \updownarrow  \vec{\tilde{\tau}}^{1}\,.   \quad 
\end{equation}
The numbers in the diagram determine the four massless families before the electroweak break, presented in 
Table~\ref{Table I.} of  appendix~\ref{cliffordfamilies}. All the scalar fields 
are in the adjoint representation of the charge groups, they are either triplets or singlets. 

Let us denote the scalar fields 
-- triplets and singlets -- which in the electroweak 
break gain  nonzero vacuum expectation values (all in pairs $(\mp)$), with a  common vector
\begin{eqnarray}
\label{ninescalars}
\Phi^{Ai}& \equiv& \Phi^{Ai}_{\mp}= (\vec{\tilde{A}}^{\tilde{N}_L}_{\mp}\,, \,\vec{\tilde{A}}^{1}_{\mp}\,, \,
\vec{A}^{1}_{\mp}\,,\vec{A}^{o}_{\mp}\,,A^{Y}_{\mp}\,,
\, A^{Y'}_{\mp})\,, \nonumber\\
\Phi^{Ai}_{\mp}&=& (\Phi^{Ai}_{7} \pm i \Phi^{Ai}_{8})\,,\quad A=\{\tilde{N}_L,\tilde{1}, 1, o, Y,Y'\}\,,
\end{eqnarray}
and let us assume that the effective potential $V(\Phi^{Ai})  $, coupling all the (assumed to be real) 
scalar fields, is renormalizable 
\begin{eqnarray}
\label{veff}
V(\Phi^{Ai})   &=& \sum_{A,i}\{\, -\frac{1}{2} \,  (m_{Ai})^2 (\Phi^{Ai})^2 + \frac{1}{4}\, \sum_{ B,
j}\, \lambda^{Ai\,Bj}\, (\Phi^{Ai})^2 \, (\Phi^{Bj})^2\}\,.
\end{eqnarray}
Couplings among the scalar fields are here chosen to be symmetric: $\lambda^{Ai\,Bj}=\lambda^{Bj\, Ai}$. 
The scalar triplets $\vec{A}^{1}_{\mp}\,$  and 
$\vec{A}^{o}_{\mp}\,$   
couple to the gauge bosons 
according to the Lagrange function  ${\mathcal L}_s$
\begin{eqnarray}
\label{scalarLagrange0}
{\mathcal L}_s &=& \sum_{A,i}\,\,(p_{0m} \Phi^{Ai}_{})^{\dagger}(p_{0}{}^{m}\, \Phi^{Ai}_{}) - V(\Phi^{Ai})\,,
\nonumber\\
p_{0m} &=& p_{m} - \{ g^{Y}\ \,Y\,A^{Y}_{m} +  g^{1}\, \vec{T}^1\, \vec{A}^{1}_{m} \}\,.
\end{eqnarray}

All the components of $\vec{\tilde{A}}^{1}_{\mp}$ and 
$\vec{\tilde{A}}^{\tilde{N}_L}_{\mp}$ and also of $A^{Q}_{\mp}$ and $Z^{Q'}_{\mp}$,   
\begin{eqnarray}
\label{scalarrelated}
A^{Y}_{\mp}&=& -\sin \vartheta_{1} Z^{Q'}_{\mp} + \cos \vartheta_{1}\,A^{Q}_{\mp}\,,\nonumber\\ 
A^{13}_{\mp}&=& \cos \vartheta_{1} Z^{Q'}_{\mp} + \sin \vartheta_{1}\,A^{Q}_{\mp}\,, 
\end{eqnarray}
are assumed to have after the electroweak break  nonzero 
vacuum expectation values, except the triplets $\vec{A}^{1}_{\mp}$ and $\vec{A}^{o}_{\mp}$, 
for which we assume, in order that the electromagnetic field stay massless, that only the components 
$A^{11}_{\mp}$  and $A^{o3}_{\mp}$ have  nonzero vacuum expectation values~(Eq.(\ref{tripletvacuu})), 
$<\vec{A}^{1}_{\mp}>= (v_{11\,\mp}, 0,0)\,$  and $<\vec{A}^{o}_{\mp}>= (0,v_{o\,3\,\mp}, 0)\,,$  
so that $Q\, 
<\vec{A}^{1}_{\mp}>= 0 $
and  $Q \, <\vec{A}^{o}_{\mp}>=0$. This choice relates 
the vacuum expectation values of $Z^{Q'}_{\mp}$ and $A^{Q}_{\mp}$ ($v_{Q'\,\mp}\,$  and $v_{Q\,\mp}$ 
respectively)
\begin{eqnarray}
\label{vzvarelated}
\frac{v_{Q\,\mp}}{v_{Q' \,\mp}}&=&- \tan^{-1} \vartheta_{1}\,,\quad 
\frac{v_{Y \,\mp}}{v_{Q' \,\mp}}=- \sin^{-1} \vartheta_{1}\,. 
\end{eqnarray}
\section{Superposition of scalar fields in minimization procedure on tree level}
\label{minimization}

Let us look for the minimum of the  potential of Eq.~(\ref{veff}) and search for the 
mass eigen states 
on the tree level.  First we find the first 
derivatives with respect to all the scalar fields and put them equal to zero
\begin{eqnarray}
\label{scalarminimum}
\frac{\partial V(\Phi^{Ai})}{\partial \Phi^{Ai}} &=& 0 = \Phi^{Ai}\,[- (m_{Ai})^2 + 
\lambda^{Ai} (\Phi^{Ai})^2  + 
\sum_{B, j}\, \lambda^{Ai\,Bj} \,(\Phi^{Bj})^2]\,.
\end{eqnarray}
Here the notation $\lambda^{Ai}: = \lambda^{AiAi}$ is used. When expressing the minimal 
values of the scalar fields, let as call them $v_{Ai}$, 
as functions of the parameters, Eq.~(\ref{scalarminimum}) leads to the  coupled 
equations for the same number of  unknowns $v_{Ai}= \Phi^{Ai}_{min}$ 
%
\begin{eqnarray}
\label{scalarv0}
- (m_{Ai})^2 + \sum_{B,j}\,\lambda^{Ai Bj} (v_{Bj})^2=0\,.
\end{eqnarray}
Looking for the second derivatives at the minimum determined by $v_{Ai}$ one finds
\begin{eqnarray}
\label{scalarsecderv0}
\frac{\partial^2 V(\Phi^{Ck})}{\partial \Phi^{Ai}\partial\Phi^{Bj} }|_{ v_{Ck}} &=& 
2 \lambda^{Ai Bj} v_{Ai}\,v_{Bj}\,. 
\end{eqnarray}
Let us  look for the basis  $\Phi^{\beta}$ (we should not forget the index ($\pm$)),  
\begin{eqnarray}
\label{scalarnew}
\Phi^{Ai} = \sum_{\beta}\, {\cal C}^{Ai}_{\beta}\, \Phi^{\beta}\,,
\end{eqnarray}
in which on the tree level the potential would be diagonal 
\begin{eqnarray}
\label{scalardiag}
V(\Phi^{\beta})   &=& \sum_{\beta}\{\, - \frac{1}{2} \, (m_{\beta})^2 (\Phi^{\beta})^2 + 
\frac{1}{4}\, \lambda^{\beta}\, (\Phi^{\beta})^4\}\,,
\end{eqnarray}
with $\frac{\partial V}{\partial \Phi^{\beta}}|_{v_{Ai}} =$ $  
  \sum_{Ai}\, \frac{\partial V}{\partial \Phi^{Ai}}
 \frac{\partial \Phi^{Ai}}{\partial \Phi^{\beta}}|_{v_{Ai}} = 0\,,$ 
 with $\Phi^{\beta}_{min}$ at these points called 
$v_{\beta}$ $=\sum_{Ai}\, {\cal C}^{Ai\, T}_{\beta}\, v_{Ai}$ ($T$ denotes transposition), 
and correspondingly with  $\frac{\partial^2 V}{\partial (\Phi^{\beta})^2}|_{ v_{\beta}}$
= $- (m_{\beta})^2 + 3 \lambda^{\beta }\, (\Phi^{\beta})^2|_{ v_{\beta}}
$= $2 \,\lambda^{\beta} \,(v_{\beta})^2 =$ $\sum_{A,i,B,k}\, 2 \lambda^{Ai Bj} \,v_{Ai}\,v_{Bj}\,
{\cal C}^{Ai}_{\beta}\,{\cal C}^{Bj}_{\beta}$.
This means that the new basis can be found by diagonalizing the matrix of the second derivatives 
at the minimum and correspondingly put to zero the determinant
\begin{eqnarray}
\label{scalardiag1}
\det \left(\begin{array}{cccc}
2 \lambda^{\tilde{N}_{L} 1}\, \, (v_{\tilde{N}_{L} 1})^2 - 2\,(m_{\beta})^2,&
2 \lambda^{\tilde{N}_{L} 1 \, \tilde{N}_{L} 2}\,v_{\tilde{N}_{L} 1}\, v_{\tilde{N}_{L} 2},&
2 \lambda^{\tilde{N}_{L} 1 \, \tilde{N}_{L} 3}\,v_{\tilde{N}_{L} 1}\, v_{\tilde{N}_{L} 3},&\ldots\\
2 \lambda^{\tilde{N}_{L} 2 \, \tilde{N}_{L} 1}\,v_{\tilde{N}_{L} 2}\, v_{\tilde{N}_{L} 1},&
2 \lambda^{\tilde{N}_{L} 2}\, \, (v_{\tilde{N}_{L} 2})^2-2\, (m_{\beta})^2,&
2 \lambda^{\tilde{N}_{L} 2 \, \tilde{N}_{L} 3}\,v_{\tilde{N}_{L} 1}\, v_{\tilde{N}_{L} 3}, &\ldots\\
\vdots&\vdots&\ddots
\end{array}\right)\,.
\end{eqnarray}
The same number of orthogonal scalar fields $\Phi^{\beta}$, with nonzero 
vacuum expectation values and nonzero masses, as we started with, follow.
To each of them one eigen value $2 (m_{\beta})^2$ corresponds, determined by the parameters
$m_{Ai}$ and $\lambda^{AiBj}$ of Eq.~(\ref{veff}).

For the time evolution of the free scalar fields one correspondingly finds for each $\beta$
\begin{eqnarray}
\label{scalaralphatimeevol}
\Phi^{\beta}(t) =  e^{-i m_{\beta} (t-t_0)}\, \Phi^{\beta}(t_0)\,.  
\end{eqnarray}
\subsection{A simple example}
\label{example2x2}

Let us examine a simple case,  one triplet, say $\vec{\tilde{A}}^{1}$, and let us call these 
three scalar states 
$\Phi^{i}$. 
Following Eq.~(\ref{scalarminimum}) one obtains 
$- (m_{i})^2 + \sum_{j}\,\lambda^{i j} (v_{j})^2=0\,,\quad {\rm for\;\;each\;\;} \,i=1,2,3$.  
Let us further simplify the example by the assumption that one of these three fields is decoupled: 
$\lambda^{i3}=0$, for $i=(1,2)$. 
Then it follows  for the vacuum expectation values $v_{i}, i\in \{1,2,3\}$
\begin{eqnarray}
\label{scalarv02}
(v_{1})^2 &=& \frac{- \lambda^{12} (m_2)^2 + \lambda^{^2} (m_1)^2}{\lambda^1 \lambda^2- (\lambda^{12})^2}\,,\quad
(v_{2})^2  = \frac{- \lambda^{12} (m_1)^2 + \lambda^{^1} (m_2)^2}{\lambda^1 \lambda^2- (\lambda^{12})^2}\,,\quad
(v_{3})^2  = \frac {(m_3)^2}{\lambda^3}\,.
\end{eqnarray}
The second derivatives at the minimum, 
$\frac{\partial^2 V(\Phi^{k})}{\partial \Phi^{i}\partial\Phi^{j} }|_{\Phi^{k}= v_{k}} = 
2 \lambda^{i j} v_{i}\,v_{j}\,$,  
lead to the determinant~(Eq.(\ref{scalardiag1})), 
from where one obtains the eigen masses
\begin{eqnarray}
\label{scalardiagres}
(m^{1,2})^{2}&=& \frac{1}{2}\,\{[\lambda^1 (v_1)^2 + \lambda^2 (v_2)^2] \mp 
\sqrt{[\lambda^2 (v_2)^2 - \lambda^1 (v_1)^2]^2+4 (\lambda^{12})^2 \,(v_1)^2\, (v_2)^2}\}\,,
\end{eqnarray}
and $(m^3)^2 = (m_{3})^2$. 
If the coupling between the two scalar components is zero, the trivial case of three uncoupled  
scalar fields  follows. In the case that the two masses, $m_{1}$ and  $m_{2}$,  are equal and that also 
the two self strengths are the same, $\lambda^1=\lambda^2$, then $(v_1)^2 = (v_2)^2 $  and the 
two eigen values for masses are  $(m^{1,2})^{2}=(v_1)^2 [(\lambda^1- \lambda^{12}), (\lambda^1+ 
\lambda^{12}] $. In the case that $\lambda^1$ and $\lambda^{12}$ are close to each other, 
the two eigen values differ a lot. 
In the case of $\lambda^{12}=0$ the two scalars would manifest as only 
one. 

Such a simplified situation illustrates that the mass eigen states  of the scalar fields might 
differ a lot from the superposition of the scalar fields which couples to any of the family members of any 
of the families, the tree level mass matrices of which are presented in 
Table~\ref{Table VIII.} and in Eq.~(\ref{adiag}) and discussed in next section~\ref{fermionmassmatrix}.

\section{Coupling of family members to  scalar fields}
\label{fermionmassmatrix}

The tree level contributions of $\vec{\tilde{A}}^{\tilde{N}_{L}}_{\mp}$  and 
$\vec{\tilde{A}}^{1}_{\mp}$ (Eq.~(\ref{factionI})) to the mass matrix of any family member 
look~\cite{NF,AN} as it is presented in Table~\ref{Table VIII.}. The notation 
$\tilde{a}^{\tilde{A}i}_{\mp}=$ $-\tilde{g}^{\tilde{A}}\, 
\tilde{v}_{\tilde{A}i\,\mp}$ is used, where $\tilde{v}_{\tilde{A}i\,\mp}$ are the vacuum expectation 
values of the corresponding scalars. Let us repeat that $\tilde{a}^{\tilde{A}i}_{\mp}$ 
distinguish among ($u^{i}\,, \nu^{i}$) ($(-)$) and ($d^{i}\,, e^{i}$) ($(+)$).
Since $\tau^{1i}$ on the right handed spinors give zero, the triplets $\vec{A}^{1}_{\mp}$, 
as well as $\vec{A}^{o}_{\mp}$
do not contribute to the fermion masses.
The contributions of  $ g^{Y}_{1}\,\cos \vartheta_{1}\,
Q\,A^{Q}_{\mp}\,, $ $g^{1}_{1}\,\cos \vartheta_{1}\, Q'\, Z^{Q'}_{\mp}\,,$ and $g^{Y'}\,Y'\,
A^{Y'}_{\mp}$ are not presented in Table~\ref{Table VIII.}. They  are  different  for each of the 
family member $\alpha =(u^i, d^i, \nu^i, e^i)$ and the same for all the families ($i = (1,2,3,4)$) 
\begin{eqnarray}
\label{adiag}
a^{\alpha}_{\mp}&=&-\{ g^{Q}_{1}\,Q^{\alpha}\,v_{Q \,\mp} + g^{Q'}_{1}\,Q^{'\alpha}\,v_{Q'\,\mp}
+ g^{Y'}\,Y^{'\alpha}\,v_{Y'\,\mp} 
\}\,, 
\end{eqnarray}
with $Q^{\alpha}\,,$ $Q^{'\alpha}$ and $Y^{'\alpha}$, 
which  are eigen values 
of the corresponding operators for the spinors state $\alpha$.
When assuming that the triplets of Eq.~(\ref{tripletvacuu}) determine  masses of the weak bosons 
on the tree level, Eq.~(\ref{vzvarelated})  relates the vacuum expectation values  
$v_{Q\,\mp}\, $, $v_{Q'\,\mp}\,$ and $\, \tan \vartheta_{1}$. 
Correspondingly we have $a^{\alpha}_{\mp}= \frac{ g^{Y}_{1}}{\sin \vartheta_{1}}\,Y^{\alpha}\,v_{Q' \,\mp} 
- g^{Y'}\,Y^{'\alpha}\,v_{Y'\,\mp}\,$. %

Also possible Majorana term, appearing in the theory, and manifesting in higher orders,
is not in the table of the tree level contributions. 
 \begin{table}
 \begin{center}
\begin{tabular}{|r||c|c|c|c||}
\hline
 $i$&$ 1 $&$ 2 $&$ 3 $&$4 $\\
\hline\hline
$1 $&
$ - \frac{1}{2}\,( \tilde{a}^{13}_{\mp} + \tilde{a}^{\tilde{N}^{3}_{L}}_{\mp})$&
$\tilde{a}^{\tilde{N}_{L}^{-}}_{\mp}$&$0$&
$   \tilde{a}^{1-}_{\mp}$  \\
\hline
$2$ &  $ \tilde{a}^{\tilde{N}_{L}^{+}}_{\mp} $ &
$ \frac{1}{2}( -\tilde{a}^{13}_{\mp } + \tilde{a}^{\tilde{N}^{3}_{L}}_{\mp}) $&
$\tilde{a}^{1-}_{\mp}   $ &$0$\\
\hline
$3$ & $0$& $\tilde{a}^{1+}_{\mp}$&
$  \frac{1}{2}\,( \tilde{a}^{13}_{\mp}  - \tilde{a}^{\tilde{N}^{3}_{L}}_{\mp}) $ &
 $\tilde{a}^{\tilde{N}_{L}^{-}}_{\mp}$ \\
\hline
$4$ & $\tilde{a}^{1+}_{\mp}$& $0$&$  \tilde{a}^{\tilde{N}_{L}^{+}}_{\mp}  $ &
 $\frac{1}{2}\,( \tilde{a}^{13}_{\mp}  + \tilde{a}^{\tilde{N}^{3}_{L}}_{\mp}) $
\\
\hline\hline
\end{tabular}
 \end{center}
 \caption{\label{Table VIII.}  The  contributions of the fields ($-\tilde{g}^{1}\, 
 \vec{\tilde{\tau}}^{1}\,\vec{\tilde{A}}^{1}_{\mp}\,,
 \, -\tilde{g}^{\tilde{N}_{L}}\, \vec{\tilde{N}}^{i}_{L}\, \vec{\tilde{A}}^{\tilde{N}_L}_{\mp}$) 
 to the mass matrices on the tree level (${\cal M}_{(o)}$) for the lower four  families  of quarks and
 leptons after the electroweak break are presented.  
The notation $\tilde{a}^{\tilde{A}i}_{\mp}=$ $-\tilde{g}^{\tilde{A}}\, \tilde{v}^{\tilde{A}i}_{\mp}$ is used. 
 }
\end{table}
Loop corrections, to which also the massive gauge fields and dynamical massive scalar fields contribute,  
are expected to strongly  influence fermions properties. 
These calculations are in progress~\cite{AN} and look so far 
promising in offering the right answers for the masses and mixing matrices of fermions.

Let $\psi^{\alpha}_{(L,R)}$ denote massless and $\Psi^{\alpha}_{(L,R)}$  massive four vectors 
for each family member  $\alpha= (u_{L,R}, d_{L,R}, \nu_{L,R}, e_{L,R})$ after taking into account 
loop corrections in all orders~\cite{NF,AN},  
$\psi^{\alpha}_{(L,R)} = V^{\alpha}_{(L,R)} \,\Psi^{\alpha}_{(L,R)} \,$,
and let
$(\psi^{\alpha \,k}_{(L,R)}\,$,  $\,\Psi^{\alpha\, k}_{(L,R)} )$ 
be any component of the four vectors, massless and massive, respectively.
On the tree level  we have 
$\psi^{\alpha}_{(L,R)}=V^{\alpha}_{(o)}\:
\Psi^{\alpha \,(o)}_{(L,R)}$ 
and
\begin{equation}
\label{treenotation}
          < \psi^{\alpha}_{L}|\gamma^0 \, {\cal M}^{\alpha}_{(o)}\,
 |\psi^{\alpha}_{R}> = < \Psi^{\alpha \,(o)}_{L}|\gamma^0 \,V^{\alpha\,
 \dagger}_{(o)}\, {\cal M}^{\alpha}_{(o)}\,V^{\alpha}_{(o)}\,|\Psi^{\alpha}_{R \,(o)}>,
\end{equation}
with 
${\cal M}^{\alpha}_{(o)k\, k'}=\sum_{A,i}\, (-g^{Ai} \, v_{Ai\, \mp})\,\, C^{\alpha}_{k\,k'}\,$. 
The coefficients $ C^{\alpha}_{k\,k'}$ can be read from  Table~\ref{Table VIII.}. 
It then follows
\begin{eqnarray}
\label{Phipsi}
\overline{\Psi}^{\alpha}\,V^{\alpha \dagger}_{(o)}\, {\cal M}^{\alpha}_{(o)}\,V^{\alpha }_{(o)}\:
 \Psi^{\alpha} &=& \overline{\Psi}^{\alpha}\,{\rm diag}(m^{\alpha}_{(o)1}\,,\cdots\,,m^{\alpha}_{(o)4})\,
 \Psi^{\alpha}\,, \nonumber\\
V^{\alpha \dagger}_{(o)}\, {\cal M}^{\alpha}_{(o)}\,V^{\alpha }_{(o)}&=& \Phi^{\alpha}_{f(o)}\,.
\end{eqnarray}
The coupling constants $m^{\alpha}_{(o)k}$ (in some units) of the dynamical scalar fields 
$\Phi^{\alpha}_{f(o) \,k}$ to 
the family member  $\Psi^{\alpha \,k}$ belonging to the $k^{th}$ family are on the tree level 
correspondingly equal to
\begin{eqnarray}
\label{Phipsiex}
 (\Phi^{\alpha}_{\Psi(o)})_{k\,k'}\,\Psi^{\alpha\,k'} &=& \delta_{k\,k'}
 \,m^{\alpha}_{(o)k}\,\Psi^{\alpha\,k}\,. 
\end{eqnarray}

The superposition of scalar fields $(\Phi^{\alpha}_{f(o)})$, which couple to fermions~\footnote{Let me 
here refer to the simple case of subsect.~\ref{example2x2} by paying attention to the reader 
that in Table~\ref{Table VIII.} the two vacuum expectation values of each of the two scalar triplets, 
($\tilde{a}^{\tilde{N}_{L} \mp}, \tilde{a}^{\tilde{N}_{L} 3}$) and 
($\tilde{a}^{1 \mp}, \tilde{a}^{1 3}$), are expected to have the property ($\tilde{a}^{\tilde{N}_{L} + } \approx$
$\tilde{a}^{\tilde{N}_{L} -}$) and ($\tilde{a}^{1 +} \approx$ $ \tilde{a}^{1 -}$),  respectively,  
or at least very close to this. Then  superposition of the scalar fields, to 
which different families couple, might  differ a lot.} and depend 
on the quantum numbers $\alpha$ and  $k$, are in general   different 
from the superposition $\Phi^{\beta}$  (Eqs.~(\ref{scalarnew},\ref{scalaralphatimeevol})), which 
are the 
mass eigen states. Each family member $\alpha$ of each massive family $k$ couples in  general 
to different superposition of scalar fields.

The two kinds of superposition are expressible with each other
\begin{eqnarray}
\label{PhiPhiPsi}
\Phi^{\alpha }_{f(o)\,k} &=& \sum_{\beta} \, D^{\alpha\,\beta}_{ k} \,\Phi^{\beta }\,. 
\end{eqnarray}
\section{Scalar triplets  bring masses to weak bosons}
\label{scalartriplets}

According to the assumption  of Eqs.(\ref{scalarLagrange0}) and (\ref{tripletvacuu}) there are triplets 
$\vec{A}^{1}_{\mp}$  
and $\vec{A}^{o}_{\mp}$ with their nonzero vacuum expectation values, which manifest  after the 
electroweak phase transition as a replacement  for vielbeins and those spin connection fields 
which couple to the weak boson gauge field $Z^{Q'}_{m}$ and $W^{1\pm}_{m}$. 

This replacement might not be the right one, yet it is useful to  demonstrate that scalar fields 
to which the weak bosons couple might differ a lot from the mass eigen states as well as from  
those to which fermions couple.  

The operators $(-g^{1}\,\vec{T} \vec{A}^{1}_{m}- g^{Y}\,Y\,\, A^{Y}_{m})$ operating on the scalar 
triplets ($\vec{A}^{1}_{\mp}$, $\vec{A}^{o}_{\mp}$)  form the 
$3 \times 3$ matrix 
\begin{eqnarray}
\label{bosonmass0}
 \left(\begin{array}{ccc}
 -g^1\,A^{13}_m + g^Y A^{Y}_m & -g^1 \frac{1}{\sqrt{2}}\,(A^{11}_m -i A^{12}_m)& 0\\
 -g^1 \frac{1}{\sqrt{2}}\,(A^{11}_m + i A^{12}_m)   & g^Y A^{Y}_m &
 -g^1 \frac{1}{\sqrt{2}}\,(A^{11}_m -i A^{12}_m)\\
 0 &-g^1 \frac{1}{\sqrt{2}}\,(A^{11}_m + i A^{12}_m)& g^1\,A^{13}_m + g^Y A^{Y}_m
 \end{array}\right)\,.
\end{eqnarray}
Assuming a new superposition of gauge fields  as in  the {\it standard model}
\begin{eqnarray}
\label{newbosons}
A^{13}_m    = \cos \theta \,Z^{Q'}_{m} + \sin \theta\, A^{Q}_{m}\,,\;\; 
A^{Y}_m      = -\sin \theta \,Z^{Q'}_{m} + \cos \theta \, A^{Q}_{m}\,,\;\;
W^{1\,\pm}_m = \frac{1}{\sqrt{2}}(A^{11}_m \mp i A^{12}_m)\,,
\end{eqnarray}
we end up with the matrix which, when being applied on $(v_{11\,\mp},0,0)$,  leads to the three vector 
 $( -\frac{g^1}{\cos \theta}\,v_{11\,\mp}\, Z^{Q'}_{m}\,,  -g^1 \,v_{11\,\mp}\,W^{1-}_m \,, 0)$. 
 The operators   $(-g^{1}\,\vec{T} \vec{A}^{1}_{m}- g^{Y}\,Y\,\, A^{Y}_{m})$ operating on 
 the scalar triplet $(0,v_{o3 \mp},0)$ gives the three vector 
 $(v_{03\,\mp}\,W^{1+}_m, 0, v_{03\,\mp}\,W^{1-}_m )$.

The contribution of both  triplets to the masses  of the weak bosons on the tree level is correspondingly
\begin{eqnarray}
\label{bosonmasstriplet1}
  (g^1)^2 \,|v_{11\,\mp}|^2(\frac{1}{(\cos \theta)^2}\, Z^{Q'}_{m} Z^{Q'\,m} + 
  (1+ 2\,|\frac{v_{o\,3\,\mp}}{v_{11\,\mp}}|^2 ) W^{1\,+}_{m}W^{1\,- \,m} )\,.
\end{eqnarray}
%

The {\it standard model} weak bosons mass term obtained on 
a tree level by the Higgs, which is a weak doublet,  
carries the hyper charge $Y=\frac{1}{2}$ and  has the vacuum 
expectation value $(0, v)$, is equal to
%
$ (\frac{1}{2})^2\, (g^1)^2 \,v^2(\frac{1}{(\cos \theta_{W})^2}\, Z_{m}Z^{m} +  2\,W^{+}_{m}W^{- \,m} )\,$.
%
The triplet $\vec{A}^{1}_{\mp}$ 
gives for the factor of $\sqrt{2}$ too large mass of $Z^{m}$  with respect to the mass of $W^{\pm\,m}$. 
If the triplet $\vec{A}^{o}_{\mp}$  contributes so that $|\frac{g^{0}\, v_{o\,3\,\mp}}{g^1\,v_{11\,\mp}}|^2= $ 
$\frac{1}{2}$, the sum of both contributions give the weak boson mass ratio on the tree level as  the 
{\it standard model} does.

Since the Lagrange function  (Eq.~(\ref{scalarLagrange0}))  
is  assumed (from the symmetry requirements), not derived from the 
starting Lagrange function for the vielbeins and the two kinds of the spin connection fields, the 
result is not trustable, but meaningful, if the requirement that $|\frac{g^{0}\, v_{o\,3\,\mp}}{g^1\,v_{11\,\mp}}|^2= $ 
$\frac{1}{2}$ can be proved.   
One can also hardly expect that  the tree level contributions to the ratio 
of the weak bosons masses is in a good agreement with the measured ones,  
while the tree level mass matrices for fermions, when diagonalized, do not  
fit well to the experimental  masses and mixing matrices. 
Since the so far done calculations show~\cite{AN} that loop corrections might 
lead to the mass matrices which  result in the measured properties of fermions, loop corrections to 
the weak boson masses might influence the ratio of boson masses as well~\footnote{   
Let us also recognize that to come from the {\it spin-charge-family} theory to the {\it standard model} 
not only must  
all the scalar fields originated in the scalar components of the vielbeins and the two kinds of the 
spin connection fields be replaced by one scalar field, which is 
a weak doublet, but must the effect of these scalar fields on the fermion properties  be replaced by their 
measured  properties. 
}.

Let us conclude this section by looking at the time evolution of the  two scalar triplets $\vec{A}^{k}_{\mp}$,
$k=1,o$
(Eqs.~(\ref{scalarnew}, \ref{scalaralphatimeevol}))
\begin{eqnarray}
\label{timeevolutionAi}
\Phi^{ki}(t,t_0) = 
\sum_{\beta}\, {\cal C}^{ki}_{\beta}\, e^{-i m_{\beta} (t-t_0)} \, \Phi^{\beta}\,, \quad k=1,o\,.
\end{eqnarray}
\section{Conclusions}
\label{discussionsandconclusions}

It is demonstrated (on the tree level only) that according to the {\it spin-charge-family} theory -- 
which predicts the families of fermions and their charges, the gauge fields and several scalar fields --  
each family member ($\alpha = (u,d,\nu,e)$) $\Psi^{\alpha \, k}$ of each family ($k=(1,2,3,4)$) 
couples to a different superposition of the scalar fields ($\Phi^{\alpha}_{f(0)}$), Eq.~(\ref{Phipsi}), 
with the coupling constant proportional to its (fermion) mass (Eq.~(\ref{Phipsiex})). Each of these 
superposition differs from the scalar fields (Eq.~(\ref{timeevolutionAi})), assumed to be  triplets 
(Eqs.~(\ref{tripletvacuu}), (\ref{scalarLagrange0})), which contribute to the masses of the weak 
gauge bosons (Eq.~(\ref{scalarLagrange0})). The scalar mass eigen states ($\Phi^{\beta}$) form the 
superposition (Eqs.(\ref{scalarnew}), (\ref{scalardiag})), which again differ from all the above 
mentioned superposition of the scalar fields. Properties depend on the parameters, the values of which 
are in this paper not discussed. 

Although the ratio of the masses of the weak gauge bosons $\frac{m_{W}}{m_{Z}}$, determined by the 
 assumed (not derived from the starting action) triplets  on the tree level, does  agree with 
 the {\it standard model} prediction ($\frac{m_{W}}{m_{Z}}= \frac{\cos \theta_{1}}{\sqrt{2}}$)
 on the tree level under the 
 condition that the ratio ($\frac{v_{o\,3\,\mp}}{ v_{11\,\mp}}|^2=\frac{1}{2}$, 
 (Eqs.~(\ref{bosonmasstriplet1})),  
the loop corrections might 
drastically change the results for 
fermions~\cite{AN}, weak bosons and scalars properties. 
Yet the 
analyses clearly shows, that several scalar fields can hardly be seen in all the experiments as only 
one Higgs as predicted by the {\it standard model}. 
 
It appears as a great challenge to explain, if the {\it standard model} is really a low energy 
effective manifestation of the {\it spin-charge-family} theory (or of any other theory which is able 
to explain the existence of the families and correspondingly of several scalar fields leading to 
mass matrices and Yukawa couplings), why the {\it standard model}, with the scalar fields replaced  
by a weak doublet (with the charges in the fundamental representation), does predict the mass ratio  
already on the tree level in such a good agreement with the experimental data (although with the 
experimentally obtained fermion properties which take into account to  
some extent all loop corrections). 
Since taking into account loop corrections in all orders manifests a complicated many body problem, 
such explanation might be very difficult as we learn from several very efficient effective 
theories in many body problems.

Let me conclude with the predictions of the {\it spin-charge-family} theory on the tree level: 
Observations of the scalar fields at  the LHC and other experiments might differ  a lot from 
the predictions of the {\it standard model}, although so far the experimental data have shown 
no disagreement with the {\it standard model} predictions. 
A systematic study of predictions of the {\it spin-charge-family} theory  is needed. 
The predictions for the observation of the scalar fields is in progress~\cite{GNinprogress}.

\appendix*

\section{The technique for representing spinors~\cite{NF,norma,norma93,hn0203}}
\label{cliffordfamilies} 

The technique~\cite{NF,norma93,hn0203,norma} can be used to construct a spinor basis for any dimension $d$
and any signature in an easy and transparent way. Equipped with the graphic presentation of basic states,  
the technique offers an elegant way to see all the quantum numbers of states with respect to the two 
Lorentz groups with the infinitesimal generators of the groups $S^{ab}$ and $\tilde{S}^{ab}$, 
as well as transformation properties of the states under any Clifford algebra object $\gamma^a$ 
and $\tilde{\gamma}^a$, $ \{ \gamma^a, \gamma^b\}_{+} = 2\eta^{ab}\,$,    
$\{ \tilde{\gamma}^a, \tilde{\gamma}^b\}_{+}= 2\eta^{ab}\,$, 
$\{ \gamma^a, \tilde{\gamma}^b\}_{+} = 0\,$, for any $d$, even or odd.  

Since the Clifford algebra objects $S^{ab}= (i/4) (\gamma^a \gamma^b - \gamma^b \gamma^a)\,$ and 
$\tilde{S}^{ab}= (i/4) (\tilde{\gamma}^a \tilde{\gamma}^b 
- \tilde{\gamma}^b \tilde{\gamma}^a)\,$ close the algebra of the Lorentz group, while 
$\{S^{ab}, \tilde{S}^{cd}\}_{-}= 0\,$, 
$S^{ab}$ and $\tilde{S}^{ab}$ form the equivalent representations to each other.
If $S^{ab}$ are used to determine spinor representations in $d$ dimensional space, and after the 
break of symmetries, the spin and the charges in $d= (1+3)$, can $\tilde{S}^{ab}$ be used to describe 
families of spinors. 

To make the technique simple  the graphic presentation of nilpotents and projectors was introduced~\cite{hn0203}. 
For even $d$ we have
\begin{eqnarray}
\stackrel{ab}{(k)}:&=& 
\frac{1}{2}(\gamma^a + \frac{\eta^{aa}}{ik} \gamma^b)\,,\quad \quad
\stackrel{ab}{[k]}:=
\frac{1}{2}(1+ \frac{i}{k} \gamma^a \gamma^b)\,,
\label{signature}
\end{eqnarray}
with the properties $k^2 = \eta^{aa} \eta^{bb}$ and 
\begin{eqnarray}
        S^{ab}\, \stackrel{ab}{(k)}= \frac{1}{2}\,k\, \stackrel{ab}{(k)}\,,\quad  
        S^{ab}\, \stackrel{ab}{[k]}= \frac{1}{2}\,k \,\stackrel{ab}{[k]}\,,\quad 
\tilde{S}^{ab}\, \stackrel{ab}{(k)}= \frac{1}{2}\,k \,\stackrel{ab}{(k)}\,,\quad  
\tilde{S}^{ab}\, \stackrel{ab}{[k]}=-\frac{1}{2}\,k \,\stackrel{ab}{[k]}\,.
\label{grapheigen}
\end{eqnarray}
One recognizes 
that $\gamma^a$ transform  $\stackrel{ab}{(k)}$ into  $\stackrel{ab}{[-k]}$, never to 
$\stackrel{ab}{[k]}$, while $\tilde{\gamma}^a$ transform  $\stackrel{ab}{(k)}$ into 
$\stackrel{ab}{[k]}$, never to $\stackrel{ab}{[-k]}$ 
\begin{eqnarray}
&&\gamma^a \stackrel{ab}{(k)}= \eta^{aa}\stackrel{ab}{[-k]},\; 
\gamma^b \stackrel{ab}{(k)}= -ik \stackrel{ab}{[-k]}, \; 
\gamma^a \stackrel{ab}{[k]}= \stackrel{ab}{(-k)},\; 
\gamma^b \stackrel{ab}{[k]}= -ik \eta^{aa} \stackrel{ab}{(-k)}\,,\nonumber\\
&&\tilde{\gamma^a} \stackrel{ab}{(k)} = - i\eta^{aa}\stackrel{ab}{[k]},\;
\tilde{\gamma^b} \stackrel{ab}{(k)} =  - k \stackrel{ab}{[k]}, \;
\tilde{\gamma^a} \stackrel{ab}{[k]} =  \;\;i\stackrel{ab}{(k)},\; 
\tilde{\gamma^b} \stackrel{ab}{[k]} =  -k \eta^{aa} \stackrel{ab}{(k)}\,. 
\label{snmb:gammatildegamma}
\end{eqnarray}

Let us add some useful relations
\begin{eqnarray}
\stackrel{ab}{(k)}\stackrel{ab}{(k)}& =& 0\,, \quad \quad \stackrel{ab}{(k)}\stackrel{ab}{(-k)}
= \eta^{aa}  \stackrel{ab}{[k]}\,, \quad  
\stackrel{ab}{[k]}\stackrel{ab}{[k]} = \stackrel{ab}{[k]}\,, \quad \quad
\stackrel{ab}{[k]}\stackrel{ab}{[-k]}= 0\,, 
 \nonumber\\
\stackrel{ab}{(k)}\stackrel{ab}{[k]}& =& 0\,,\quad \quad \quad \stackrel{ab}{[k]}\stackrel{ab}{(k)}
=  \stackrel{ab}{(k)}\,, \quad \quad \quad \stackrel{ab}{(-k)}\stackrel{ab}{[k]}=
 \stackrel{ab}{(-k)}\,,
\quad \quad \stackrel{ab}{[k]}\stackrel{ab}{(-k)} =0 \,.  
\label{graphbinoms}
\end{eqnarray}
Defining
\begin{eqnarray}
\stackrel{ab}{\tilde{(\pm i)}} = 
\frac{1}{2} \, (\tilde{\gamma}^a \mp \tilde{\gamma}^b)\,, \quad
\stackrel{ab}{\tilde{(\pm 1)}} = 
\frac{1}{2} \, (\tilde{\gamma}^a \pm i\tilde{\gamma}^b)\,, 
\label{deftildefun}
\end{eqnarray}
it follows 
\begin{eqnarray}
\stackrel{ab}{\tilde{( k)}} \, \stackrel{ab}{(k)}& =& 0\,, 
\quad \;
\stackrel{ab}{\tilde{(-k)}} \, \stackrel{ab}{(k)} = -i \eta^{aa}\,  \stackrel{ab}{[k]}\,,
\quad\;
\stackrel{ab}{\tilde{( k)}} \, \stackrel{ab}{[k]} = i\, \stackrel{ab}{(k)}\,,
\quad\;
\stackrel{ab}{\tilde{( k)}}\, \stackrel{ab}{[-k]} = 0\,.
\label{graphbinomsfamilies}
\end{eqnarray}
We define the vacuum $|\psi_0>$ so that 
$< \;\stackrel{ab}{(k)}^{\dagger} \stackrel{ab}{(k)}\; > = 1\,$ and $< \;\stackrel{ab}{[k]}^{\dagger}
 \stackrel{ab}{[k]}\; > = 1\,$.

Making a choice of the Cartan subalgebra set of the algebra $S^{ab}$ and $\tilde{S}^{ab}$  
$(S^{03}, S^{12}, S^{56}, S^{78}, \cdots)$ and $(\tilde{S}^{03}, \tilde{S}^{12}, \tilde{S}^{56}, 
\tilde{S}^{78},\cdots )$ 
an eigen state of all the members of the Cartan  subalgebra, representing a weak chargeless  
$u_{R}$-quark with spin up, hyper charge ($2/3$) and  colour ($1/2\,,1/(2\sqrt{3})$), for example, 
can be written as 
%
$ \stackrel{03}{(+i)}\stackrel{12}{(+)}|\stackrel{56}{(+)}\stackrel{78}{(+)}
||\stackrel{9 \;10}{(+)}\stackrel{11\;12}{(-)}\stackrel{13\;14}{(-)} |\psi \rangle$  $= 
\frac{1}{2^7} 
(\gamma^0 -\gamma^3)(\gamma^1 +i \gamma^2)$ $| (\gamma^5 + i\gamma^6)(\gamma^7 +i \gamma^8)$ $||
(\gamma^9 +i\gamma^{10})(\gamma^{11} -i \gamma^{12})(\gamma^{13}-i\gamma^{14})
|\psi \rangle \,$.
%
This state is an eigen state of all $S^{ab}$ and $\tilde{S}^{ab}$ which are members of the Cartan 
subalgebra. The definition of the  charges can be found in the ref.~\cite{pikanorma,NF}.

In Table~\ref{Table I.} the eightplet of quarks of a particular colour charge ($\tau^{33}=1/2$, 
 $\tau^{38}=1/(2\sqrt{3})$) and the $U(1)_{II}$ charge ($\tau^{4}=1/6$) is presented in our 
 technique~\cite{norma93,hn0203}, as products of nilpotents and projectors. 
 \begin{table}
 \begin{center}
 \begin{tabular}{|r|c||c||c|c|c|r|r|}
 \hline
 i&$$&$|^a\psi_i>$&$\Gamma^{(1,3)}$&$ S^{12}$&
 $\tau^{13}$&$Y$&$Q$\\
 \hline\hline
 && ${\rm Octet} \,\,{\rm of \; quarks}$&&&&&\\
 \hline\hline
 1&$ u_{R}^{c1}$&$ \stackrel{03}{(+i)}\,\stackrel{12}{(+)}|
 \stackrel{56}{(+)}\,\stackrel{78}{(+)}
 ||\stackrel{9 \;10}{(+)}\;\;\stackrel{11\;12}{[-]}\;\;\stackrel{13\;14}{[-]} $
 &1&$\frac{1}{2}$&0&$\frac{2}{3}$&$\frac{2}{3}$\\
 \hline 
 2&$u_{R}^{c1}$&$\stackrel{03}{[-i]}\,\stackrel{12}{[-]}|\stackrel{56}{(+)}\,\stackrel{78}{(+)}
 ||\stackrel{9 \;10}{(+)}\;\;\stackrel{11\;12}{[-]}\;\;\stackrel{13\;14}{[-]}$
 &1&$-\frac{1}{2}$&0&$\frac{2}{3}$&$\frac{2}{3}$\\
 \hline
 3&$d_{R}^{c1}$&$\stackrel{03}{(+i)}\,\stackrel{12}{(+)}|\stackrel{56}{[-]}\,\stackrel{78}{[-]}
 ||\stackrel{9 \;10}{(+)}\;\;\stackrel{11\;12}{[-]}\;\;\stackrel{13\;14}{[-]}$
 &1&$\frac{1}{2}$&0&$-\frac{1}{3}$&$-\frac{1}{3}$\\
 \hline 
 4&$ d_{R}^{c1} $&$\stackrel{03}{[-i]}\,\stackrel{12}{[-]}|
 \stackrel{56}{[-]}\,\stackrel{78}{[-]}
 ||\stackrel{9 \;10}{(+)}\;\;\stackrel{11\;12}{[-]}\;\;\stackrel{13\;14}{[-]} $
 &1&$-\frac{1}{2}$&0&$-\frac{1}{3}$&$-\frac{1}{3}$\\
 \hline
 5&$d_{L}^{c1}$&$\stackrel{03}{[-i]}\,\stackrel{12}{(+)}|\stackrel{56}{[-]}\,\stackrel{78}{(+)}
 ||\stackrel{9 \;10}{(+)}\;\;\stackrel{11\;12}{[-]}\;\;\stackrel{13\;14}{[-]}$
 &-1&$\frac{1}{2}$&$-\frac{1}{2}$&$\frac{1}{6}$&$-\frac{1}{3}$\\
 \hline
 6&$d_{L}^{c1} $&$\stackrel{03}{(+i)}\,\stackrel{12}{[-]}|
 \stackrel{56}{[-]}\,\stackrel{78}{(+)}
 ||\stackrel{9 \;10}{(+)}\;\;\stackrel{11\;12}{[-]}\;\;\stackrel{13\;14}{[-]} $
 &-1&$-\frac{1}{2}$&$-\frac{1}{2}$&$\frac{1}{6}$&$-\frac{1}{3}$\\
 \hline
 7&$ u_{L}^{c1}$&$\stackrel{03}{[-i]}\,\stackrel{12}{(+)}|
 \stackrel{56}{(+)}\,\stackrel{78}{[-]}
 ||\stackrel{9 \;10}{(+)}\;\;\stackrel{11\;12}{[-]}\;\;\stackrel{13\;14}{[-]}$
 &-1&$\frac{1}{2}$&$\frac{1}{2}$&$\frac{1}{6}$&$\frac{2}{3}$\\
 \hline
 8&$u_{L}^{c1}$&$\stackrel{03}{(+i)}\,\stackrel{12}{[-]}|\stackrel{56}{(+)}\,\stackrel{78}{[-]}
 ||\stackrel{9 \;10}{(+)}\;\;\stackrel{11\;12}{[-]}\;\;\stackrel{13\;14}{[-]}$
 &-1&$-\frac{1}{2}$&$\frac{1}{2}$&$\frac{1}{6}$&$\frac{2}{3}$\\
 \hline\hline
 \end{tabular}
 \end{center}
 \caption{\label{Table I.} The 8-plet of quarks - the members of $SO(1,7)$ subgroup of the 
 group $SO(1,13)$, belonging to one Weyl 
 spinor representation of  $SO(1,13)$ is presented in the technique~\cite{hn0203}. 
 It contains the left handed weak charged quarks and the right handed weak chargeless quarks 
 of a particular 
 colour $(1/2,1/(2\sqrt{3}))$. Here  $\Gamma^{(1,3)}$ defines the handedness in $(1+3)$ space, 
 $ S^{12}$ defines the ordinary spin, 
 $\tau^{13}$ defines the third component of the weak charge,  
 $Y$ is the hyper charge,  
 $Q= Y + \tau^{13}$ is the 
 electromagnetic charge. The vacuum state $|\psi_0>$, on which the nilpotents and 
 projectors operate, is not shown. The basis is the massless one. One easily sees that 
 $\gamma^0 \,\stackrel{78}{(-)}$ transforms the first line ($u_{R}^{c1}$) into the seventh one ($u_{L}^{c1}$), 
 and $\gamma^0 \,\stackrel{78}{(+)}$ transforms the third line ($d_{R}^{c1}$) into the fifth one ($d_{L}^{c1}$).
 }
 \end{table}
The operators $ \tilde{S}^{ab}$ 
generate families from the starting $u_R$ quark, transforming $u_R$ quark from Table~(\ref{Table VIII.}) 
to the $u_R$ of another family,  keeping all the properties with respect to $S^{ab}$ unchanged. 
The eight families of the first 
 member of the eightplet of quarks from Table~\ref{Table I.}, for example, that is of the right 
 handed $u^{c1}_{R}$-quark  with spin $\frac{1}{2}$,  are presented in the left column of 
 Table~\ref{Table III.}. 
 The eight-plet of  the corresponding 
 right handed neutrinos with spin up is  presented  
 in the right column of the same table. All the other members of any of the eight families of quarks or 
 leptons follow  from any member of a particular family by the application of  the 
 operators  $S^{ab}$ 
 on this particular member.  
 \begin{table}
 \begin{center}
 \begin{tabular}{|r||c||c||c||c||}
 \hline
 $I_R$ & $u_{R}^{c1}$&
  $ \stackrel{03}{(+i)}\,\stackrel{12}{[+]}|\stackrel{56}{[+]}\,\stackrel{78}{(+)}||
  \stackrel{9 \;10}{(+)}\;\;\stackrel{11\;12}{[-]}\;\;\stackrel{13\;14}{[-]}$ & 
  $\nu_{R}$&
  $ \stackrel{03}{[+i]}\,\stackrel{12}{(+)}|\stackrel{56}{[+]}\,\stackrel{78}{(+)}|| 
  \stackrel{9 \;10}{(+)}\;\;\stackrel{11\;12}{(+)}\;\;\stackrel{13\;14}{(+)}$ 
 \\
 \hline
  $II_R$ & $u_{R}^{c1}$&
  $ \stackrel{03}{[+i]}\,\stackrel{12}{(+)}|\stackrel{56}{[+]}\,\stackrel{78}{(+)}||
  \stackrel{9 \;10}{(+)}\;\;\stackrel{11\;12}{[-]}\;\;\stackrel{13\;14}{[-]}$ & 
  $\nu_{R}$&
  $ \stackrel{03}{(+i)}\,\stackrel{12}{[+]}|\stackrel{56}{(+)}\,\stackrel{78}{[+]}||
  \stackrel{9 \;10}{(+)}\;\;\stackrel{11\;12}{(+)}\;\;\stackrel{13\;14}{(+)}$ 
 \\
 \hline
 $III_R$ & $u_{R}^{c1}$&
 $ \stackrel{03}{(+i)}\,\stackrel{12}{[+]}|\stackrel{56}{(+)}\,\stackrel{78}{[+]}||
 \stackrel{9 \;10}{(+)}\;\;\stackrel{11\;12}{[-]}\;\;\stackrel{13\;14}{[-]}$ & 
 $\nu_{R}$&
 $ \stackrel{03}{(+i)}\,\stackrel{12}{[+]}|\stackrel{56}{[+]}\,\stackrel{78}{(+)}||
 \stackrel{9 \;10}{(+)}\;\;\stackrel{11\;12}{(+)}\;\;\stackrel{13\;14}{(+)}$ 
 \\
 \hline
 $IV_R$ & $u_{R}^{c1}$&
  $ \stackrel{03}{[+i]}\,\stackrel{12}{(+)}|\stackrel{56}{(+)}\,\stackrel{78}{[+]}||
  \stackrel{9 \;10}{(+)}\:\; \stackrel{11\;12}{[-]}\;\;\stackrel{13\;14}{[-]}$ & 
  $\nu_{R}$&
  $ \stackrel{03}{[+i]}\,\stackrel{12}{(+)}|\stackrel{56}{(+)}\,\stackrel{78}{[+]}||
  \stackrel{9 \;10}{(+)}\;\;\stackrel{11\;12}{(+)}\;\;\stackrel{13\;14}{(+)}$ 
 \\
 \hline 
 \end{tabular}
 \end{center}
 \caption{\label{Table III.} Four  families of the right handed $u^{c1}_R$ quark with spin 
 $\frac{1}{2}$, the colour charge (${}^{c1}$ $=(\tau^{33}=1/2$, $\tau^{38}=1/(2\sqrt{3}))$, and of 
 the colourless right handed neutrino $\nu_R$ of spin $\frac{1}{2}$ are presented in the 
 left and in the right column, respectively. All the families follow from the starting one by the 
 application of the operators  $\tilde{S}^{ab}$, $a,b \in\{0,1,2,\cdots, 8\}$. 
The  generators $S^{ab}$, $a,b \in\{0,1,2,\cdots, 8\}$  transform 
equivalently the right handed   neutrino $\nu_R$ of  spin $\frac{1}{2}$ to all the colourless 
members of the same family. 
}
 \end{table}
%

%

\end{document}